\documentclass[a4paper,11pt]{article}
\pdfoutput=1 

\usepackage{jheppub}

\usepackage{graphicx, epstopdf}

\usepackage[utf8]{inputenc}

\usepackage{amsmath, amsthm, amsfonts, amssymb}
\usepackage{mathtools, mathabx, bm, bbm, scalerel, xfrac}
\usepackage{empheq, physics, slashed, cancel, braket}
\usepackage[usenames,dvipsnames]{xcolor}

\usepackage{hyperref}
\usepackage{color}

\allowdisplaybreaks

\title{Quark Mass Effects in Higgs Production}

\author[a]{Micha\l{} Czakon,}
\author[a]{Felix Eschment,}
\author[b]{Marco Niggetiedt,}
\author[c]{Rene Poncelet,}
\author[a]{Tom Schellenberger}

\affiliation[a]{Institut f\"ur Theoretische Teilchenphysik und Kosmologie, RWTH Aachen University,\\ 52056 Aachen, Germany}
\affiliation[b]{Max-Planck-Institut f\"ur Physik,\\ Boltzmannstra{\ss}e 8, 85748 Garching, Germany.}
\affiliation[c]{The Henryk Niewodnicza\'{n}ski Institute of Nuclear Physics,\\ ul.\ Radzikowskiego 152, 31-342 Krakow, Poland}

\emailAdd{mczakon@physik.rwth-aachen.de}
\emailAdd{felix.eschment@rwth-aachen.de}
\emailAdd{marco.niggetiedt@mpp.mpg.de}
\emailAdd{rene.poncelet@ifj.edu.pl}
\emailAdd{tom.schellenberger@rwth-aachen.de}

\abstract{We examine the effect of finite top- and bottom-quark masses on the Higgs production cross section in the gluon-gluon fusion channel. We employ both $\overline{\text{MS}}$ and on-shell renormalisation for the quark masses and provide a thorough comparison. Furthermore, we explore alternative treatments of quark masses, in particular in the four-flavour scheme, and investigate their impact on the cross section. Our work also presents novel predictions for differential cross sections in the Higgs rapidity. The results lead to a significant reduction of scale uncertainties, and our analysis enables us to offer well-grounded recommendations for future research in this area.}
\preprint{P3H-24-048, TTK-24-27, IFJPAN-IV-2024-9, MPP-2024-137}

\begin{document}
\maketitle
\flushbottom

\section{Introduction}
The groundbreaking discovery of the Higgs boson in 2012~\cite{ATLAS:2012yve, CMS:2012qbp} led to an ongoing effort to scrutinise whether this newfound particle behaves as predicted by the Standard Model (SM). At the Large Hadron Collider (LHC), the Higgs boson is primarily produced in gluon-gluon fusion, where two gluons produce the Higgs via a quark-triangle loop. The gluon-gluon-fusion cross section is therefore one of the central observables in Higgs-physics phenomenology. Currently, its most precise experimental determination \cite{CMS:2022dwd} is
\begin{equation}
\sigma_{ggH} = 47.1 \pm 3.8 \ \mathrm{pb}
\end{equation}
at a center of mass energy of $13\ \mathrm{TeV}$, and theoretical predictions with uncertainties negligible in comparison are desirable.

Since the top quark is by far the heaviest quark in the SM, it also has the largest impact on the cross section, hence Higgs couplings to lighter quarks are often neglected. Moreover, the top quark's large mass enables the use of the \textit{heavy-top limit} (HTL) approximation, where the top quark is treated as infinitely massive. This method is valid up to power corrections of order $m_H^2 /  m_t^2$, and significantly streamlines computations by reducing the loop order by one and removing the top-quark mass as a scale. These simplifications have allowed for calculations at next-to-leading~\cite{Dawson:1990zj,Djouadi:1991tka}, next-to-next-to-leading~\cite{Ravindran:2002dc, Harlander:2002wh, Anastasiou:2002yz}, and next-to-next-to-next-to-leading~\cite{Anastasiou:2015vya,Mistlberger:2018etf, Anastasiou:2016cez} order in $\alpha_s$ (NLO, NNLO, $\text{N}^3$LO), yielding state-of-the-art predictions with a theoretical uncertainty of approximately $6\%$. Studies have shown that the gluon-gluon-fusion channel receives particularly large perturbative corrections. In fact, the cross section is more than doubled going from LO to NLO, and NNLO corrections remain substantial, contributing roughly 20\% to the total cross section. Because of the large size of the corrections, accounting for the previously omitted finite top-quark mass is strongly advisable also at higher perturbative orders. In \textit{Higgs effective field theory} (HEFT), finite-top-quark-mass effects are included by rescaling the HTL cross section to recover the LO result with finite top-quark mass:
\begin{equation}
\sigma^{\text{N}^n\text{LO}}_{\text{HEFT}} = \frac{\sigma^{\text{LO}}}{\sigma^{\text{LO}}_{\text{HTL}}} \sigma_{\text{HTL}}^{\text{N}^n\text{LO}} \approx 1.065 \times \sigma^{\text{N}^n\text{LO}}_{\text{HTL}}.
\label{eq:HEFT_rescaling}
\end{equation}
This approach assumes that the ratio $\sigma / \sigma_{\text{HTL}}$ remains relatively constant across perturbative orders, providing a significant improvement over HTL by capturing much of the finite-top-quark-mass effects for fully inclusive observables. Nevertheless, with the theoretical precision of the N${}^3$LO predictions, this simple rescaling method is insufficient, and a more comprehensive treatment of the top-quark mass is indispensable. 

Besides to the top quark, the Higgs boson can also couple to other quark flavours in the loop. Gluon-gluon-fusion amplitudes in which the Higgs couples to light-quark flavours are suppressed by powers of $m_q^2/m_H^2$. The leading contribution of these amplitudes to the cross section arises via interference with top-quark induced gluon-gluon fusion. Despite the strong power suppression, the interference receives large logarithmic enhancements of the form $\log^2(m_q^2/m_H^2)$, rendering the total effect of lighter quark flavours on the cross section quite sizeable. 

The effect of finite top and bottom masses on the total Higgs production cross section has been studied up to NNLO in refs.~\cite{Czakon:2021yub, Czakon:2023kqm, Niggetiedt:2024nmp}, thereby reducing the previous theory uncertainty of 6\% \cite{Anastasiou:2016cez} by roughly a third. In the present study we aim to expand on this work and provide a more in-depth analysis. In particular, we want to address the questions left open in ref.~\cite{Czakon:2023kqm} and investigate the impact of different mass-renormalisation and \textit{flavour schemes} (FS). 
The treatment of heavy quarks is not unambiguous. A theory with $n_l$ light quarks taken as massless is called $n_l$-flavour scheme, $n_l$FS. In ref.~\cite{Czakon:2023kqm}, the calculation was performed in the 5FS, ergo the bottom quark was treated as massless. For a non-vanishing contribution, the bottom quark must still couple to the Higgs. Hence, it is considered massive inside closed quark loops that couple to the Higgs. Although this method ensures that the resulting amplitude is gauge invariant, the retrospective introduction of mass requires stronger justification. We want to address this issue and explore the impact on the cross section of a consistent bottom-quark treatment as a massive particle, i.e.\ the impact of working in the 4FS.

At NLO, the predictions for the top-bottom-interference contribution differ by approximately 30\% between the OS and $\overline{\text{MS}}$ schemes, raising concerns about the reliability of cross-section predictions at this order. Furthermore, in the OS scheme, the NNLO contribution to top-bottom interference is almost of the same size as the NLO correction but with opposite sign~\cite{Czakon:2023kqm}. Alternating corrections of similar size typically hint at poor perturbative convergence. Combined with the fact that the scale uncertainties increase going from NLO to NNLO, an alternative computational framework is desirable. An alternative mass-renormalisation scheme could mitigate the poor convergence and yield more reliable predictions.

Finite-quark-mass effects can also be leveraged to determine the Yukawa couplings to the respective quarks. The sensitivity of the measurements can be significantly enhanced by additionally considering differential cross sections, e.g.~the Higgs-$p_T$~\cite{Bishara:2016jga, Bonner:2016sdg} or the Higgs-rapidity~\cite{Soreq:2016rae} distribution. This is particularly useful for lighter quarks like the charm quark. CMS studies~\cite{CMS:2018gwt, CMS:2023gjz} have found an upper bound for the charm Yukawa coupling of about five times its SM value, by fitting NLO predictions to the measured distributions. Meanwhile, NNLO predictions for $p_T$ distributions with finite top and bottom mass have become available~\cite{Grazzini:2013mca, Buschmann:2014sia, Melnikov:2016emg, Lindert:2017pky,Caola:2018zye, Bonciani:2022jmb} and could potentially yield improved bounds. Rapidity distributions on the other hand are even more challenging, as they require a contribution from the three-loop virtual corrections absent in the $p_T$ distribution. Our study aims to bridge this gap, potentially enhancing the sensitivity for future analyses of this kind.

This paper is organised as follows: In section~\ref{sec:sec2} we explain our computational framework. We put a special emphasis on the modifications required for different mass-renormalisation and flavour schemes. In section~\ref{sec:sec3} we present the results for the fully-inclusive cross sections as well as differential distributions for the Higgs-$p_T$ and Higgs-rapidity. In section~\ref{sec:sec4} we give our conclusions and recommendations for future phenomenological studies. Finally, appendix~\ref{sec:appendix1} contains a comparison of our results to existing Higgs+jet analyses.

\section{Computational Details}
\label{sec:sec2}
The general workflow mirrors that of ref.~\cite{Czakon:2023kqm}. In essence, we calculate the gluon-gluon-fusion amplitudes for both real and virtual corrections, i.e.~the amplitudes for three-loop double-virtual, two-loop real-virtual and one-loop double-real corrections, and then perform the phase-space integration using Monte Carlo techniques. Soft and collinear divergences which cancel between the different-multiplicity phase spaces as well as those absorbed in the initial state renormalisation are handled with \texttt{Stripper} -- the C++ implementation of the sector-improved residue subtraction scheme \cite{Czakon:2010td,Czakon:2014oma,Czakon:2019tmo}. 

Amplitudes for double-real radiation have been calculated in ref.~\cite{Budge:2020oyl} and we use the implementation available in \texttt{MCFM} \cite{Campbell:2019dru, Campbell:1999ah, Campbell:2011bn}, which in turn uses \texttt{QCDLoop} \cite{Ellis:2007qk, Carrazza:2016gav} for the evaluation of scalar integrals. The three-loop double-virtual corrections are known in terms of large-mass and/or high-energy expansions to very high precision for both the single-quark~\cite{Czakon:2020vql} as well as the mixed-flavour contributions~\cite{Niggetiedt:2023uyk}.

The two-loop real-virtual corrections can be categorised into mixed- and single-flavour diagrams. The former consist of diagrams with two different quark flavours (see fig.\ \ref{fig:RV_4fs}) and they can be written as a product of simple one-loop integrals which we solve analytically. The single-flavour diagrams on the other hand involve genuine two-loop integrals, which we solve numerically. To do this, the amplitude is first projected to form factors and the required set of master integrals is determined by numerically solving the system of integration-by-parts identities. All appearing loop integrals are reduced to a predefined basis of master integrals with the public software \texttt{Kira}$\oplus$\texttt{FireFly} \cite{Maierhofer:2017gsa, Maierhofer:2018gpa, Klappert:2020nbg}. The reduction to master integrals is computationally very expensive, mostly because of the tremendous complexity in intermediate expressions, which is why we employ finite field methods for the interpolation of multivariate rational expressions as implemented in \texttt{FireFly} \cite{Klappert:2019emp, Klappert:2020aqs}. Even still, a full reconstruction in terms of all three dimensionless kinematic variables needed for the $gg \longrightarrow g H$, $q \bar{q} \longrightarrow g H$, $q g \longrightarrow q H$ and $\bar{q} g \longrightarrow \bar{q} H$ processes, which for example can be chosen as 
\begin{equation}
z = 1 - \frac{m_H^2}{\hat{s}}, \quad \lambda = \frac{\hat{t}}{\hat{t} + \hat{u}}, \quad \text{and} \quad x = \frac{m_q^2}{\hat{s}},
\end{equation}
with $\hat{s}, \hat{t}$ and $\hat{u}$ the usual partonic Mandelstam variables, remains challenging. Since the computation time needed for the reduction of the amplitudes strongly scales with the polynomial degree of the coefficients in front of the master integrals, it is advantageous to search for a basis in which the coefficients become small in size. We heuristically explore different choices of bases at fixed values for the kinematic variables and perform a full reduction to our favored basis of master integrals retaining the complete symbolic dependence. 

\begin{figure}
\centering
\includegraphics[scale=1.6]{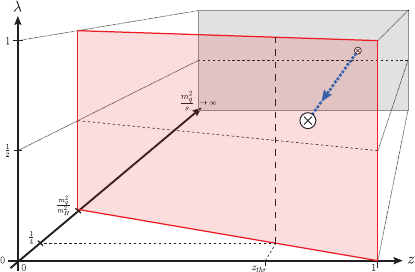}
\caption{Illustration of the strategy for numerically solving the differential equations of the master integrals. Integrals on the boundary at $x \rightarrow \infty$ are computed with the large-mass expansion. The differential equation is then solved in $x$ for multiple $z$ and a fixed value of $\lambda$ close to the symmetry axis. The red plane indicates the physical plane, with a fixed quark mass.}
\label{fig:integration_strategy}
\end{figure}

Furthermore, we use a fixed ratio of the quark and Higgs mass square, $m_q^2/m_H^2$, to eliminate one of the scales in favour of a rational number, which further simplifies the rational-function coefficients in front of the master integrals. In fact, we compute the master integrals and the complete amplitudes for five different bottom-quark masses and four different top-quark masses, listed in tab.~\ref{tab:bottom_masses}. After the reduction to master integrals, we derive a set of coupled differential equations. The integrals for the boundary condition at $x \rightarrow \infty$ can be solved with the help of the large-mass expansion. We then transport the solution to physical values of $x$, by solving the differential equations numerically using the Bulirsch-Stoer algorithm (see fig.~\ref{fig:integration_strategy}).
This is done multiple times for many values of $z$ and a fixed value of $\lambda$ close to the symmetry axis of $\lambda = 1/2$. Subsequently, we solve the differential equations in $\lambda$ to map out the entire physical plane, marked red in fig.~\ref{fig:integration_strategy}. We thus obtained three-dimensional grid points of the amplitude which we use to construct cubic spline fits during the Monte Carlo integration. To assess the accuracy of the interpolation method, we removed specific grid points---in this case we removed all respective points corresponding to $\overline{m}_b(m_H), \overline{m}_{b}(m_H/2), \overline{m}_b(m_H/4)$ or $m_t$---and reran the Monte Carlo simulation. The discrepancies between the recalculated cross sections and the original ones were below the Monte Carlo uncertainties, confirming the robustness of our interpolation method. We note that we only calculated the interference of all bottom-quark amplitudes with the OS-renormalised, as well as with the $\overline{\mathrm{MS}}$-renormalised top-quark amplitudes at the same renormalisation scale, to construct grids for cubic-spline interpolation.

To further improve the quality of the spline fits, we subtract all kinematic singularities arising from soft or collinear phase space regions from the amplitudes and add them back only after performing the interpolation. The subtraction of kinematic singularities can be simplified by first subtracting the amplitude square of the rescaled HTL (rHTL) defined by
\begin{equation}
\braket{\mathcal{M}^{(0)}_{\text{rHTL}}| \mathcal{M}^{(1)}_{\text{rHTL}}} \equiv \frac{\sigma^{\text{LO}}_i}{\sigma^{\text{LO}}_\text{HTL}} \braket{\mathcal{M}^{(0)}_{\text{HTL}}|\mathcal{M}^{(1)}_{\text{HTL}}}.
\end{equation}
$\braket{\mathcal{M}^{(0)}| \mathcal{M}^{(1)}}$ denotes the usual spin- and color-averaged interference of the LO and NLO amplitudes. $\sigma_i$ refers to either the cross section with finite top-quark mass, in which case the rHTL is identical to HEFT, or the top-bottom interference contribution. The regulated NLO contribution of the squared matrix element are given by
\begin{equation}
\begin{split}
&\braket{\mathcal{M}^{(0)}_{gg\rightarrow gH} | \mathcal{M}^{(1)}_{gg \rightarrow gH}} \vert_\text{regulated} \equiv  \braket{\mathcal{M}^{(0)}_{gg\rightarrow gH} | \mathcal{M}^{(1)}_{gg \rightarrow gH}} -  \braket{\mathcal{M}^{(0)}_{gg\rightarrow gH, \text{rHTL}} | \mathcal{M}^{(1)}_{gg \rightarrow gH, \text{rHTL}}} \\
& \quad + \braket{\mathcal{M}^{(0)}_{gg \rightarrow H} | \mathcal{M}^{(1)}_{gg \rightarrow H} - \mathcal{M}^{(1)}_{gg \rightarrow H, \text{rHTL}}}  \times \begin{cases} \frac{8 \pi \alpha_s}{\hat{t}} \braket{P_{gg}^{(0)} \!\left(\frac{\hat{s}}{\hat{s} + \hat{u}} \right)} \quad &\text{if } |\hat{t}| < |\hat{u}| \\ 
\frac{8 \pi \alpha_s}{\hat{u}} \braket{P_{gg}^{(0)} \!\left(\frac{\hat{s}}{\hat{s} + \hat{t}} \right)} \quad &\text{if } |\hat{u}| < |\hat{t}|
\end{cases}, \\[0.2cm]
&\braket{\mathcal{M}^{(0)}_{qg\rightarrow qH} | \mathcal{M}^{(1)}_{qg \rightarrow qH}} \vert_\text{regulated} \equiv  \braket{\mathcal{M}^{(0)}_{qg\rightarrow qH} | \mathcal{M}^{(1)}_{qg \rightarrow qH}} -  \braket{\mathcal{M}^{(0)}_{qg\rightarrow qH, \text{rHTL}} | \mathcal{M}^{(1)}_{qg \rightarrow qH, \text{rHTL}}} \\
& \quad - \braket{\mathcal{M}^{(0)}_{gg \rightarrow H} | \mathcal{M}^{(1)}_{gg \rightarrow H} - \mathcal{M}^{(1)}_{gg \rightarrow H, \text{rHTL}}}  \times 
\frac{8 \pi \alpha_s}{\hat{t}} \braket{P_{q\bar{q}}^{(0)} \!\left(\frac{\hat{s}}{\hat{s} + \hat{u}} \right)}, \\[0.2cm] 
&\braket{\mathcal{M}^{(0)}_{q\bar{q}\rightarrow gH} | \mathcal{M}^{(1)}_{q \bar{q} \rightarrow gH}} \vert_\text{regulated} \equiv  \braket{\mathcal{M}^{(0)}_{q \bar{q}\rightarrow gH} | \mathcal{M}^{(1)}_{q\bar{q} \rightarrow gH}} -  \braket{\mathcal{M}^{(0)}_{q \bar{q} \rightarrow gH, \text{rHTL}} | \mathcal{M}^{(1)}_{q \bar{q} \rightarrow gH, \text{rHTL}}}, \\
\end{split} 
\label{eq:IR_regulation}
\end{equation} 
where the averaged splitting functions are defined as
\begin{equation}
 \braket{P_{gg}^{(0)} \!\left(z \right)} = 2 C_A \left( \frac{z}{1 - z} + \frac{1 - z}{z} + z (1 - z) \right) \text{ and } \braket{P_{q\bar{q}}^{(0)}(z)} = T_F \left( 1 - 2 z (1 - z) \right).
\label{eq:rHTL}
\end{equation}
By subtracting the one-loop squared amplitude in the rHTL $\braket{\mathcal{M}^{(0)}_{i\rightarrow f, \text{rHTL}} | \mathcal{M}^{(1)}_{i \rightarrow f, \text{rHTL}}}$, we already capture most of the singularities, namely all singularities arising from one-loop soft functions or collinear splitting functions. These would only act on the difference between the LO form factor and its rHTL counter part, which vanishes by design (see eq.~\eqref{eq:rHTL}). The only singularities which are not yet removed stem from either a soft and collinear splitting of the NLO form factors. These singularities are then subtracted with the corresponding tree-level splitting function (see second and fourth line of eq.~\eqref{eq:IR_regulation}). Note that the splitting function also removes the soft divergences. The regulated squared matrix element is now free of non-integrable kinematic singularities. 

\subsection{$\overline{\text{MS}}$ scheme}
Compared to the calculation in the OS scheme, for the $\overline{\text{MS}}$ scheme we need to introduce two modifications: first, the running of the mass has to be considered, and second, the renormalisation of the amplitudes has to be altered. The latter does not require us to redo the entire calculation from scratch, in fact the $\overline{\text{MS}}$-renormalised amplitudes can be easily derived from the OS-renomalised amplitudes. To do this we first use the renormalisation-scheme invariance of the bare quark mass,
\begin{equation}
Z_m^{\overline{\text{MS}}}(\mu) \overline{m}(\mu) = Z_m^{\text{OS}} m^{\text{OS}},
\end{equation}
to derive the relation between the OS- and the $\overline{\text{MS}}$-renormalised masses:
\begin{equation}
m^{\text{OS}}  = \overline{m}(\mu) \left(1 + c_1(\mu) \frac{\alpha_s(\mu)}{\pi} + c_2(\mu) \left( \frac{\alpha_s(\mu)}{\pi} \right)^2 + \mathcal{O}(\alpha^3)\right).
\label{eq:mOS2mMS}
\end{equation}
The mass-renormalisation constants were first derived in refs.~\cite{Gray:1990yh, TARRACH1981384}. The renormalisation procedure must also be consistent with the used FS, that means in the 5FS, the bottom quark has to be massless inside closed bottom loops even when renormalising the bottom mass. Similarly, we have to decouple the top quark, and in the 4FS also the bottom quark, from the running of $\alpha_s$:
\begin{equation}
\begin{split}
&\alpha_s^{(n_l)} (\mu) = \zeta \left(\alpha_s^{(n_l + 1)}(\mu), \ln \frac{m^2}{\mu^2} \right) \alpha_s^{(n_l + 1)} (\mu),  \\
& \zeta(\alpha_s, L) = 1 + \frac{\alpha_s}{4 \pi} \frac{4}{3} T_F L + \left( \frac{\alpha_s}{4 \pi} \right)^2 \left[ -\frac{14}{3} + T_F \left( \frac{20}{3} C_A + 4 C_F \right) L + \frac{16}{9} T_F^2 L^2 \right].
\end{split}
\end{equation}
Here $C_F = 4/3,\ C_A = 3,\ T_F = 1/2$ and $n_l$ is the number of light quarks.
We then use Eq.~\eqref{eq:mOS2mMS} recursively to eliminate any residual logarithmic dependence on the OS mass and finally obtain
\begin{equation}
\begin{split}
c_1(&\mu) = C_F \left(1 -\frac{3}{4} \ln\! \left(\frac{\overline{m}(\mu)^2}{\mu^2} \right) \right), \\ 
c_2(&\mu) = -\frac{1}{384} C_F \bigg(C_A \left(-132 \ln^2\!\left(\frac{\overline{m}(\mu)^2}{\mu^2} \right) +740 \ln\! \left(\frac{\overline{m}(\mu)^2}{\mu^2} \right) +144 \zeta (3)-1111+\pi ^2 \left( 32-96 \ln (2) \right) \right) \\ 
& \hspace{1cm} +3 C_F \bigg(-36 \ln^2\! \left(\frac{\overline{m}(\mu)^2}{\mu^2} \right) -36 \ln\! \left(\frac{\overline{m}(\mu)^2}{\mu^2} \right) -96 \zeta (3)+71+8 \pi ^2 (8 \ln (2)-5)\bigg) \\ 
& \hspace{1cm} +4 n_l T_F\left(12 \ln^2\! \left( \frac{\overline{m}(\mu)^2}{\mu^2} \right) -52 \ln\! \left( \frac{\overline{m}(\mu)^2}{\mu^2} \right) +8 \pi ^2+71\right) \bigg) \\ 
& \hspace{-0.1cm}  -  \sum_i \frac{1}{96} C_F T_F \bigg(12 \ln^2\! \left( \frac{\overline{m}(\mu)^2}{\mu^2} \right) -24 \ln\! \left(\frac{\overline{m}(\mu)^2}{\mu^2} \right) \ln\! \left( \frac{\overline{m}_i(\mu)^2}{\mu^2} \right) -52 \ln\! \left(\frac{\overline{m}(\mu)^2}{\mu^2} \right) +32 \ln\!\left( \frac{\overline{m}_i(\mu)^2}{\mu^2} \right) \\ 
& \hspace{1cm} +48 (x_i-1)^2 \left(x_i^2+x_i+1\right) \text{Li}_2(1-x_i)-48
\left(x_i^4+x_i^3+x_i+1\right) \left(\text{Li}_2(-x_i) + \ln\!(x_i) \ln\!(x_i + 1) \right) \\ 
& \hspace{1cm} + 48 x_i^4 \ln ^2(x_i) + 48 x_i^2 \ln (x_i) - 16 \pi^2 (x_i^3 + x_i) +72 x_i^2 +71\bigg),
\end{split}
\end{equation}
where we defined $ x_i = \frac{\overline{m}_i(\mu)}{\overline{m}(\mu)}$.
The sum runs over all heavy-quark flavours which are decoupled from the running of $\alpha_s$. This includes $\overline{m}$ itself in case the corresponding quark is decoupled. Subsequently, the amplitudes can simply be expanded to find the relation between the two schemes: 
\begin{equation}
\begin{split}
    & \mathcal{M}^{\overline{\text{MS}}}(\overline{m}(\mu)) = \mathcal{M}^{\text{OS}}(\overline{m}(\mu)) + \delta \mathcal{M} (\overline{m}(\mu)), \\ 
    & \delta \mathcal{M}^{(1)} (\overline{m}(\mu)) = \overline{m}(\mu)c_1(\mu) \frac{\alpha_s(\mu)}{\pi}\frac{\mathrm{d} \mathcal{M}^{\text{OS},(0)}}{\mathrm{d} m} \bigg \vert_{m = \overline{m}(\mu)} , \\
    & \delta \mathcal{M}^{(2)} (\overline{m}(\mu)) = \overline{m}(\mu) \left[ c_1(\mu) \frac{\alpha_s(\mu)}{\pi} \frac{\mathrm{d} \mathcal{M}^{\text{OS},(1)}}{\mathrm{d} m}\bigg \vert_{m = \overline{m}(\mu)}  + c_2(\mu) \left(\frac{\alpha_s(\mu)}{\pi}\right)^2 \frac{\mathrm{d}\mathcal{M}^{\text{OS},(0)}}{\mathrm{d}m}\bigg \vert_{m = \overline{m}(\mu)}  \right] \\ 
    & \hspace{8cm}+ \frac{1}{2}\left(\overline{m}(\mu) c_1(\mu) \frac{\alpha_s(\mu)}{\pi} \right)^2 \frac{\mathrm{d}^2 \mathcal{M}^{\text{OS},(0)}}{\mathrm{d} m^2} \bigg \vert_{m = \overline{m}(\mu)} ,
\end{split}
\end{equation}
where the superscript indicates the perturbative order in $\alpha_s$.

To compute the running of the quark masses, we use the public software package \texttt{CRunDec}~\cite{Schmidt:2012az} at four-loop accuracy.

\begin{table}
\centering
\caption{Mass ratios squared, $m_b^2/m_H^2$ and $m_t^2/m_H^2$, used for the generation of numerical values of the two-loop four-point amplitudes. We chose values corresponding to the OS masses, according to PDG recommendations \cite{Workman:2022ynf}, and $\overline{\text{MS}}$-masses at scales relevant for 7-point variation w.r.t.\ to a central scale of $m_H/2$. For the latter, we used $\overline{m}_b(\overline{m}_b) = 4.18\ \text{GeV}$, as recommended by the PDG and $\overline{m}_t(\overline{m}_t) = 162.7\ \mathrm{GeV}$. The dynamic scale \eqref{eq:scale} used for our differential cross-section predictions can exceed $m_H$, and reaches values of up to $1$~TeV. At this high scale, the bottom-quark mass is $\overline{m}_b(1\ \mathrm{TeV}) = 2.41~\mathrm{GeV}$. To avoid extrapolations outside our numerical grids for the amplitudes, we therefore included another mass value, $\overline{m}_b^{\text{min}}$, below this minimum.}
\begin{tabular}{cccc}
Grid mass & Value [GeV] & Approximate ratio $m_{b,t}^2/m_H^2$ & Relative error [\textperthousand] \\
\hline
$m_b$    & 4.78  & $\frac{1}{684}$ & 0.2  \\
$\overline{m}_b(m_H)$ & 2.789 & $\frac{1}{2011}$ & 0.9 \\
$\overline{m}_b(m_H/2)$ & 2.961 & $\frac{1}{1782}$ & 0.2 \\ 
$\overline{m}_b(m_H/4)$ & 3.170 & $\frac{1}{1557}$ & 1.0\\ 
$\overline{m}_b^{\text{min}}$ & 1.67 & $\frac{1}{5602}$ & 0.1 \\ 
$m_t$ & $172.4$ & $\frac{23}{12}$ & 4 \\ 
$\overline{m}_t(m_H)$ & $166.1$ & $\frac{136}{77}$ & 0.0 \\ 
$\overline{m}_t(m_H/2)$ & $176.2$ & $\frac{149}{75}$ & 0.0 \\ 
$\overline{m}_t(m_H/4)$ & $188.2$ & $\frac{213}{94}$ & 0.0
\end{tabular}
\label{tab:bottom_masses}
\end{table}

\subsection{4-flavour scheme}
\begin{figure}
\centering
\includegraphics[scale=0.6]{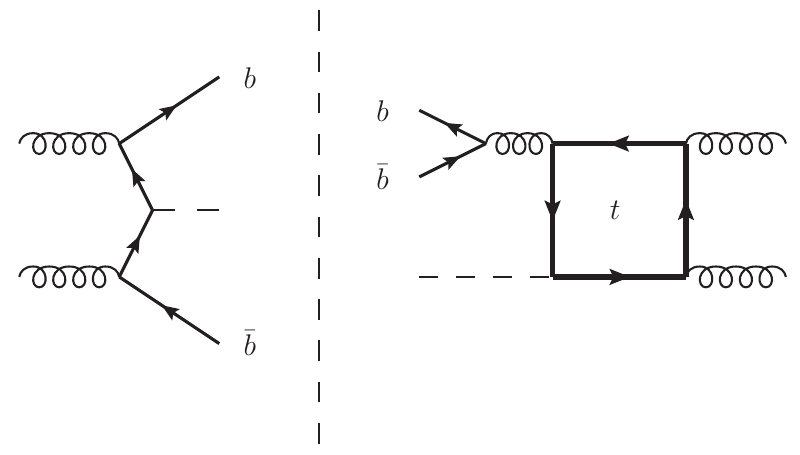}
\caption{Example interference diagram for a contribution to the bottom-bottom fusion channel at NLO in the 4FS. The contribution is proportional to $Y_b Y_t$. Because the Higgs couples to an external bottom quark, this kind of diagrams is not considered in the present calculation as a correction to gluon-gluon fusion.}
\label{fig:bbFusion}
\end{figure}

In the 5FS, the bottom quark is treated as massless except in closed fermion loops that couple to the Higgs. This treatment is formally equivalent to introducing two versions of the bottom quark: one with zero mass and another with finite mass which is absent from the proton. Since the Yukawa coupling to the Higgs is an arbitrary parameter in QCD, all infrared poles and gauge dependence must cancel independently of this parameter. This simple argument ensures that, even though we only consider a subset of all Feynman diagrams allowed in the SM, the resulting cross section will be internally consistent. Also, note that if the bottom-quark mass is discarded while maintaining a finite Yukawa coupling, the cross section will vanish all together, as the scalar nature of the coupling requires a helicity flip inside fermion loops. 
\begin{figure}
    \centering
    \includegraphics[scale=.7]{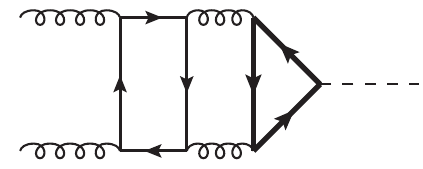}
    \caption{Example diagram contributing to the three-loop double-virtual corrections to Higgs production containing two fermion loops of different flavours. A full list of diagrams can be found in fig.~1 of ref.~\cite{Czakon:2020vql}. With the thick fermion line corresponding to a top quark and the thin line corresponding to a bottom quark in the 4-flavour scheme, this diagram contains quarks of two different masses.}
    \label{fig:formfactor_4fs}
\end{figure}
\begin{figure}
    \centering
    \includegraphics[width=.9\textwidth]{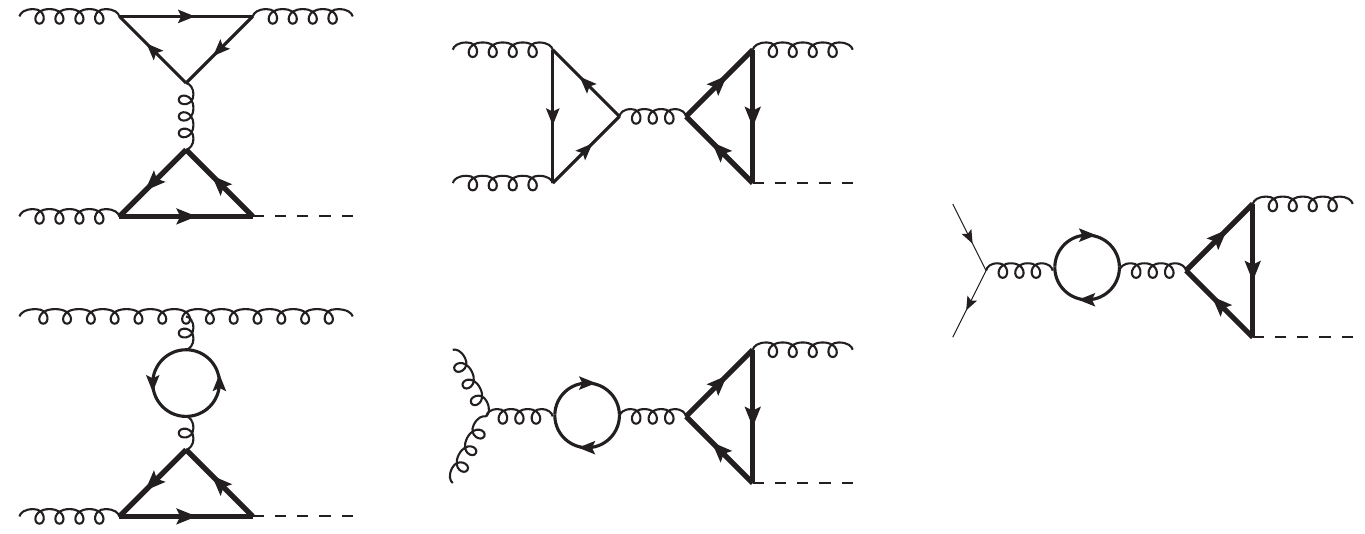}
    \caption{Diagrams with two fermion loops of different flavours contributing to the real-virtual corrections to Higgs production. The set is complete except for permutations of external gluons and reversal of the thick fermion-loop-line direction. Initial-state quarks are massless. Quark-gluon initial states can be obtained from the quark-antiquark case by crossing. Relevance to the 4-flavour scheme as in fig.~\ref{fig:formfactor_4fs}.}
    \label{fig:RV_4fs}
\end{figure}
\begin{figure}
    \centering
    \includegraphics[width=.85\textwidth]{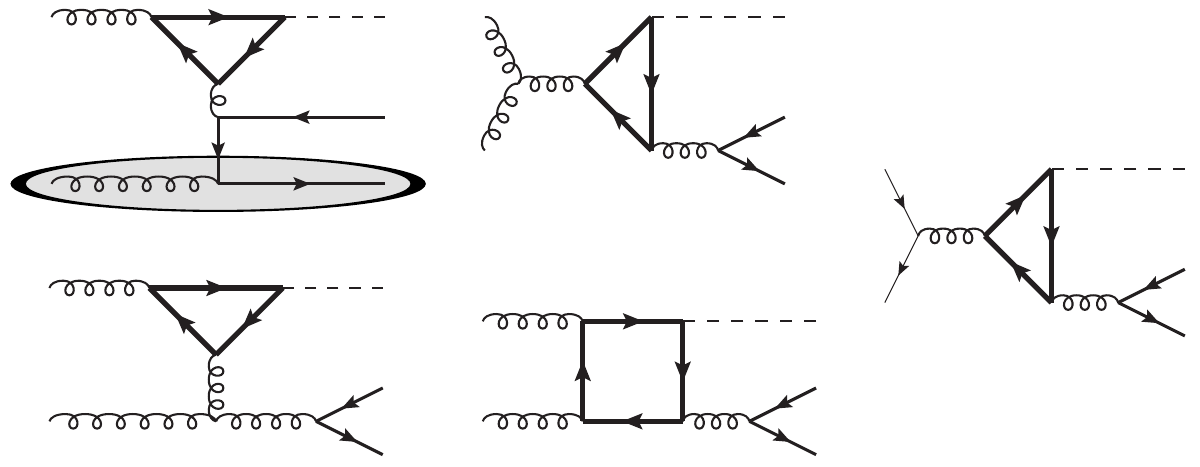}
    \caption{Diagrams corresponding to the real radiation of a quark-antiquark pair in Higgs production. Permutation of external particles and reversal of fermion-loop direction contribute additional diagrams. In the 4-flavour scheme, radiation of massive bottom quarks has to be taken into account. Infrared divergences are regulated by the finite mass and cancel with the contributions from figs.~\ref{fig:formfactor_4fs} and~\ref{fig:RV_4fs}. The shaded region in the first diagram might be shifted to the PDFs in the 5-flavour scheme, whereas bottom quarks are absent from the initial state in the 4-flavour scheme.}
    \label{fig:RR_4fs}
\end{figure}
Although the above approach is consistent in the framework of this calculation, the inhomogeneous treatment of light quark masses depending on the loop structure is rather ad hoc. It is therefore desirable to have an alternative computational framework without these contentions. For that reason we also perform the calculation in the 4FS, in which the bottom quark is always treated as a massive particle, and hence excluded from the proton PDFs. Then, amplitudes like those depicted in figs.~\ref{fig:formfactor_4fs} or \ref{fig:RV_4fs} exhibit logarithmic mass enhancements, which cancel once the radiation of a massive quark pair (see fig.~\ref{fig:RR_4fs}) is taken into account as well. These one-loop amplitudes are computed with \texttt{Recola} \cite{Actis:2016mpe, Denner:2017wsf}. Note that we still exclude diagrams in which external bottom quarks couple to the Higgs, corresponding to the tree-level amplitude $gg\longrightarrow b\bar{b} H$, as these diagrams as well as all the interference contributions are accounted for in corrections to the bottom-bottom fusion production channel \cite{Dittmaier:2003ej} in the 4FS (see fig.~\ref{fig:bbFusion}).

\begin{table*}
\caption{HEFT cross section in the 5-flavour scheme and for different bottom-quark masses in the 4-flavour scheme. In the last column the cross section and the scale variation are computed with the $\overline{\text{MS}}$-mass. The results are computed for LHC @ 13 TeV using the \texttt{NNPDF31\_nnlo\_as\_0118} PDF set in the 5FS and the \texttt{NNPDF31\_nnlo\_as\_0118\_nf\_4} PDF set in the 4FS. The central scale is chosen at $\mu_R = \mu_F = m_H/2$. The scale uncertainties are determined with seven-point variation. Results for the 5FS are computed with \texttt{SusHi}~\cite{Harlander:2012pb, Harlander:2016hcx}.}
\label{tab:HEFT_4fs}
\centering
\begin{tabular}{cccccc}
\hline
Order & \multicolumn{5}{c}{$\sigma_\text{HEFT}$ [pb]} \\
\hline
\hline
\multicolumn{6}{c}{$\sqrt{s}=13$~TeV} \\
\hline
& 5FS & 4FS  & 4FS & 4FS & 4FS  \\
& & $m_b=0.01$~GeV &  $m_b=0.1$~GeV & $m_b=4.78$~GeV & $\overline{m}_b(\overline{m}_b) = 4.18$ GeV \\
\hline
$\mathcal{O}(\alpha_s^2)$ & $+16.30$ & +16.27 & +16.27 & +16.27 & $16.27$\\
LO & $16.30^{+4.36}_{-3.10}$ & $16.27^{+4.63}_{-3.22}$ & $16.27^{+4.63}_{-3.22}$ & $16.27^{+4.63}_{-3.22}$ & $16.27^{+4.63}_{-3.22}$ \\
\hline
$\mathcal{O}(\alpha_s^3)$ & +21.14 & +20.08(3) & +20.08(3) & +20.08(3) & +20.08(3) \\
NLO & $37.44^{+8.42}_{-6.29}$ & $36.35(3)^{+8.57}_{-6.32}$ & $36.35(3)^{+8.57}_{-6.32}$ & $36.35(3)^{+8.57}_{-6.32}$ & $36.35(3)^{+8.57}_{-6.32}$ \\
\hline
$\mathcal{O}(\alpha_s^4)$ & +9.72 & +10.8(4) & +11.1(4) & +9.5(2) & $+9.6(2)$ \\
NNLO & $47.16^{+4.21}_{-4.77}$ & $47.2(4)^{+5.4}_{-5.4}$ & $47.5(4)^{+5.4}_{-5.5}$ & $45.9(2)^{+4.3}_{-4.9}$ & $46.0(2)^{+4.4}_{-5.0}$\\
\hline
\end{tabular}
\end{table*}
To validate that the cross section is free from any enhancements due to the bottom quark mass, we computed the Higgs production cross section in top-induced gluon-gluon fusion within the HEFT framework for various bottom-quark masses\footnote{We note that the rescaling factor used to define HEFT does not depend on the flavour scheme in use. As indicated in eq.~\eqref{eq:HEFT_rescaling}, only the effect of a finite top-quark mass is taken into account in the rescaling.}. The results are presented in tab.~\ref{tab:HEFT_4fs}. At LO, the only difference between the 4FS and 5FS is the used PDF set which is why there is only a marginal discrepancy between the cross sections. At NLO, the bottom quark does not appear in any 4FS amplitudes yet, hence the cross section is still independent of the bottom quark mass. In the 5FS however, the $bg \longrightarrow Hb$ channel is now opening up, causing a more significant deviation from the 4FS than at LO. Finally at NNLO the 4FS cross sections become dependent on the bottom quark mass through the diagrams depicted in fig.~\ref{fig:formfactor_4fs}, \ref{fig:RV_4fs} and \ref{fig:RR_4fs} with the top quark, represented by the thick fermion lines, integrated out, i.e.\ all top-quark loops shrink down to effective vertices. We found that the 4FS results for extremely low bottom-quark masses are numerically very similar to the 5FS result, and we see no sign of divergence. The findings also highlight the sizeable impact of finite-quark-mass effects, even if the quark does not directly couple to the Higgs, as the cross section is shifted by more than 2\% for realistic bottom-quark masses (last two columns of tab.~\ref{tab:HEFT_4fs}). This few percent effect seems to be in rough agreement with the order of magnitude of this effect in Higgs+jet production \cite{Pietrulewicz:2023dxt}.

\section{Results}
\label{sec:sec3}
In this section we present our results for the finite-quark-mass effects in the gluon-gluon-fusion channel for various computational setups. Following the Higgs Working Group recommendations \cite{LHCHiggsCrossSectionWorkingGroup:2016ypw}, our reference computation uses the 5FS, an OS-renormalised top quark with a pole mass of $m_t = 173.06$~GeV\footnote{The Higgs-Working-Group recommends using an OS-mass of $m_t = 172.4$~GeV. The different value was chosen because of the rationalisation procedure described in sec.~\ref{sec:sec2}. We have verified, that choosing a different mass has a negligible impact on the final result.} and an $\overline{\text{MS}}$-renormalised bottom mass of $\overline{m}_b(\overline{m}_b) = 4.18$~GeV. We then explore the impact of alternative mass-renormalisation schemes for both the top and the bottom quark. In addition, we perform the computation in the 4FS to examine if the flavour schemes are compatible.

We use two scale choices: for fully-inclusive cross sections we use a fixed scale with a central value of $\mu = m_H/2$, while differential cross sections are computed with a dynamic scale and a central value of 
\begin{equation}
\mu = \frac{H_T}{2} \equiv \frac{1}{2} \left(\sqrt{m_H^2 + p_{T}^2} + \sum_i p_{i, T} \right)
\label{eq:scale}
\end{equation}
where $p_{T},\ p_{i, T}$ are the transverse momenta of the Higgs and the $i$-th final state parton, respectively. In the $\overline{\mathrm{MS}}$ scheme, the mass of the bottom quark is also evaluated at $\mu$. The scale was chosen to match previous studies \cite{Lindert:2017pky, Bonciani:2022jmb, Jones:2018hbb}.

\begin{table*}[t]
\caption{Effect of the finite top-quark mass on the gluon-gluon fusion cross section for two different computational setups. The results are computed for LHC @ 13 TeV using the \texttt{NNPDF31\_nnlo\_as\_0118} PDF set in the 5FS and the \texttt{NNPDF31\_nnlo\_as\_0118\_nf\_4} PDF set in the 4FS. The central scale is chosen at $\mu_R = \mu_F = m_H/2$. The scale uncertainties are determined with seven-point variation.}
\label{tab:top-HEFT}
\centering
\begin{tabular}{ccc}
\hline
Order & \multicolumn{2}{c}{$(\sigma_{t} - \sigma_\text{HEFT})$ [pb]} \\
\hline
\hline
\multicolumn{3}{c}{$\sqrt{s}=13$~TeV} \\
\hline
& 5FS & 4FS \\
& $m_t = 173.06$ GeV &  $m_t = 173.06$ GeV \\ 
& & $\overline{m}_b(\overline{m}_b)=4.18$ GeV\\
\hline
LO & - & - \\
\hline
$\mathcal{O}(\alpha_s^3)$ & $-0.30$  &  $-0.27$ \\
NLO & $-0.30^{+0.10}_{-0.17}$ & $-0.27^{+0.09}_{-0.16}$ \\
\hline
$\mathcal{O}(\alpha_s^4)$ & $+0.14$ & $+0.12$ \\
NNLO & $-0.16^{+0.13}_{-0.03}$ & $-0.15^{+0.10}_{-0.02}$\\
\hline
\end{tabular}
\end{table*}

Tab.~\ref{tab:top-HEFT} illustrates the impact of a finite top-quark mass on the cross section, i.e.\ the difference between the result in the full theory restricted to Higgs production through a closed top-quark loop, $\sigma_t$, and in HEFT. The difference between the 5FS and 4FS (second and third column of tab.~\ref{tab:top-HEFT}) only amounts to $-0.01\ \mathrm{pb}$ at NNLO. We remind, however, that there is a significant $-2.5\%$ shift between the 5FS and 4FS results in HEFT itself (see tab.~\ref{tab:HEFT_4fs}). The influence of the flavour scheme beyond HEFT is tiny, because the effects of a finite top-quark mass are power-suppressed and at NNLO only amount to roughly 3\textperthousand \ of the total cross section.

In tab.~\ref{tab:topSchemeDifference} the difference of the cross section $\sigma_t$ in the two renormalisation schemes considered for the top-quark mass is presented. We find that it only amounts to $-0.01\ \mathrm{pb}$ at NNLO.

\begin{table}[h]
    \centering
    \caption{Difference of cross sections for Higgs production through a closed top-quark loop with the top-quark mass defined either in the $\overline{\mathrm{MS}}$ or the OS scheme. The results are computed for LHC @ 13 TeV using the \texttt{NNPDF31\_nnlo\_as\_0118} PDF set. The central scale is chosen at $\mu_R = \mu_F = m_H/2$. The scale uncertainties are determined with seven-point variation.}
    \begin{tabular}{cc}
    \hline
        Order & $(\sigma_t^{\overline{\mathrm{MS}}} - \sigma_t^{\mathrm{OS}})$ [pb] \\ 
    \hline
    \hline
    \multicolumn{2}{c}{$\sqrt{s}=13$~TeV} \\ 
    \hline
    $\mathcal{O}(\alpha_s^2)$     & $-0.04$ \\ 
    LO & $-0.04^{+0.12}_{-0.17}$ \\ 
    \hline
    $\mathcal{O}(\alpha_s^3)$ & $+0.02$ \\ 
    NLO & $-0.02^{+0.14}_{-0.30}$ \\ 
    \hline 
    $\mathcal{O}(\alpha_s^4)$ & $+0.01$ \\ 
    NNLO & $-0.01^{+0.12}_{-0.24}$ \\
    \hline
    \end{tabular}
    \label{tab:topSchemeDifference}
\end{table}

Tab.~\ref{tab:top-bottom} presents the results for the top-bottom interference contribution to the cross section $\sigma_{t\times b}$. As already noticed in ref.~\cite{Czakon:2023kqm}, the interference converges quite badly if the bottom-quark mass is OS-renormalised, as indicated by the alternating perturbative corrections of a similar order of magnitude (see third column of tab.~\ref{tab:top-bottom}). Furthermore, scale uncertainties fail to correctly estimate the effects of higher orders and actually increase from NLO to NNLO, reaching 15\%. However, results for $\overline{\text{MS}}$-renormalised bottom-quark masses (second column) show a much better perturbative convergence and additionally, we observe a decrease in scale uncertainties to below 8\%\footnote{Adapting a mixed renormalisation scheme, in which only the Yukawa coupling is renormalised in the $\overline{\text{MS}}$ scheme, while the mass in the propagators is kept in the OS scheme, also results in a better convergence and reduced scale uncertainties as noted in ref.~\cite{Czakon:2023kqm}. Without the inclusion of electroweak corrections this mixed approach is valid, but if these effects are taken into account as well, the Higgs vacuum expectation value acquires an additional finite renormalisation that propagates to the masses of gauge bosons. Since the masses of the gauge bosons are defined in the on-shell scheme by default, the mixed approach is inconsistent. Ignoring this issue, we note that the main improvements to the perturbation series of our problem are already captured by the mixed scheme.}. For the top-quark mass, the renormalisation scheme seems to have very little impact on the final result. Lastly, the NNLO result in the 4FS is 5.2\% lower than in the 5FS and agrees well within scale uncertainties. The magnitude of this decrease is of a similar size as observed in HEFT (see tab.~\ref{tab:HEFT_4fs}). 
\begin{table*}[t]
\caption{Top-bottom interference contribution to the gluon-gluon fusion cross section for various computational setups. The results are computed for LHC @ 13 TeV using the \texttt{NNPDF31\_nnlo\_as\_0118} PDF set in the 5FS and the \texttt{NNPDF31\_nnlo\_as\_0118\_nf\_4} PDF set in the 4FS. The central scale is chosen at $\mu_R = \mu_F = m_H/2$. The scale uncertainties are determined with seven-point variation.}
\label{tab:top-bottom}
\centering
\begin{tabular}{ccccc}
\hline
Order & \multicolumn{4}{c}{$\sigma_{t\times b}$ [pb]} \\
\hline
\hline
\multicolumn{5}{c}{$\sqrt{s}=13$~TeV} \\
\hline
& 5FS & 5FS  & 5FS & 4FS \\
& $m_t = 173.06$ GeV & $m_t = 173.06$ GeV &  $\overline{m}_t(\overline{m}_t) = 162.7$ GeV &  $m_t = 173.06$ GeV \\ 
& $\overline{m}_b(\overline{m}_b) = 4.18$ GeV & $m_b = 4.78$ GeV & $\overline{m}_b(\overline{m}_b) = 4.18$ GeV & $\overline{m}_b(\overline{m}_b)=4.18$ GeV\\
\hline
$\mathcal{O}(\alpha_s^2)$ & $-1.11$ & $-1.98$ & $-1.12$ & $-1.15$ \\
LO & $-1.11^{+0.28}_{-0.43}$ & $-1.98^{+0.38}_{-0.53}$  & $-1.12^{+0.28}_{-0.42}$ & $-1.15^{+0.29}_{-0.45}$\\
\hline
$\mathcal{O}(\alpha_s^3)$ & $-0.65$ & $-0.44$ & $-0.64$ & $-0.66$ \\
NLO & $-1.76^{+0.27}_{-0.28}$ & $-2.42^{+0.19}_{-0.12}$ & $-1.76^{+0.27}_{-0.28}$ & $-1.81^{+0.28}_{-0.30}$ \\
\hline
$\mathcal{O}(\alpha_s^4)$ & $+0.02$ & $+0.43$ & $-0.02$ & $-0.02$ \\
NNLO & $-1.74(2)^{+0.13}_{-0.03}$ & $-1.99(2)^{+0.29}_{-0.15}$ & $-1.78(1)^{+0.15}_{-0.03}$ & $-1.83(2)^{+0.14}_{-0.03}$\\
\hline
\end{tabular}
\end{table*}

\begin{figure}
\centering
\begin{minipage}[h]{0.49\textwidth}
\includegraphics[width=\textwidth]{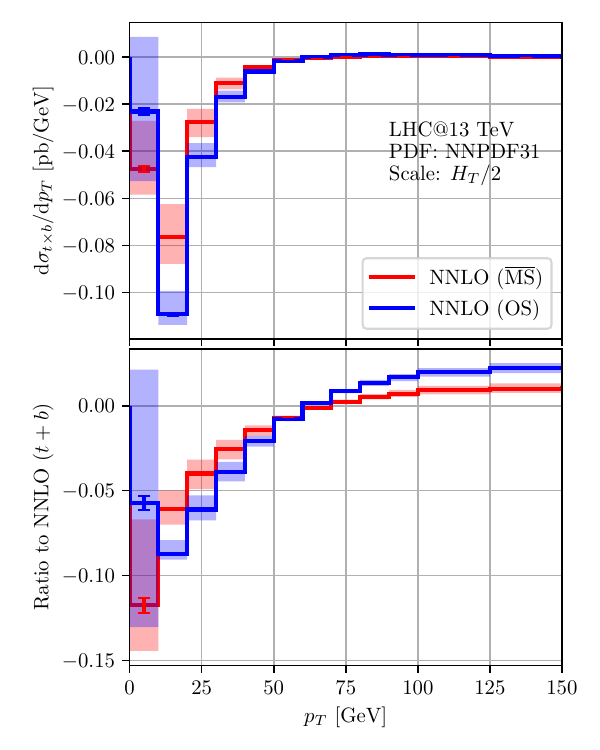}
\end{minipage}
\begin{minipage}[h]{0.49\textwidth}
\includegraphics[width=\textwidth]{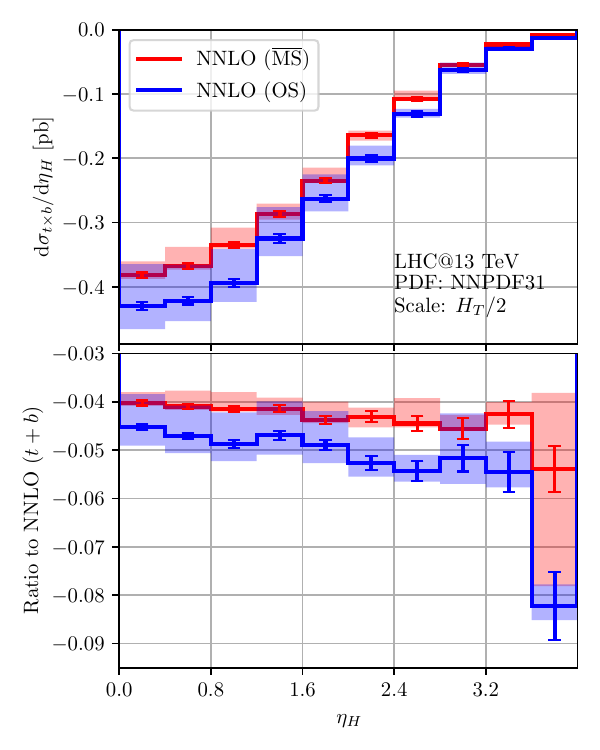}
\end{minipage}
\caption{Interference contribution to the Higgs-$p_T$ (left) and Higgs-rapidity (right) distribution. Compared are the 5FS results for an $\overline{\text{MS}}$- and OS-renormalised bottom-quark mass. The lower panel shows the ratio to the distribution including finite bottom- and top-quark-mass effects, i.e. finite top-quark masses and the interference to the bottom quark, at NNLO. Coloured bands indicate the envelope of the seven-point scale variation and the error bars represent the Monte Carlo uncertainties.}
\label{fig:pT_yH_comparison}
\end{figure}

Similar to the total cross section, the Higgs-rapidity distribution due to the top-bottom interference (right panel of fig.~\ref{fig:pT_yH_comparison}) also shows that the $\overline{\text{MS}}$ renormalisation generally yields smaller scale uncertainties. The different schemes are compatible within the estimated error bands. The Higgs-$p_T$ distribution (left panel of fig.~\ref{fig:pT_yH_comparison}) reveals that the main improvement of the uncertainties comes from the low-$p_T$ region, specifically the first bin below 10~GeV. For higher $p_T$, the scale uncertainties are of very similar size, and even slightly smaller for the OS scheme. At low $p_T$, the effect of finite bottom-quark masses is highly relevant, reaching almost 8\% of the total cross section, whereas the effect is almost negligible (below 1\%) above 50~GeV. The rapidity distribution shows less-pronounced features, closely resembling a constant shift of about $-4\%$ across all rapidity bins. The $p_T$ distributions were compared to ref.~\cite{Caola:2018zye} for the OS-renormalised cross section (see fig.~\ref{fig:Caola_comparison}). We find good agreement except for very low $p_T$, where the central value is identical but our scale uncertainties are slightly smaller. The $\overline{\text{MS}}$-renormalised $p_T$ distribution was also successfully checked against ref.~\cite{Bonciani:2022jmb} (see fig.~\ref{fig:Bonciani_comparison}).

\begin{figure}
\centering
\begin{minipage}[t]{0.49\textwidth}
\includegraphics[width=\textwidth]{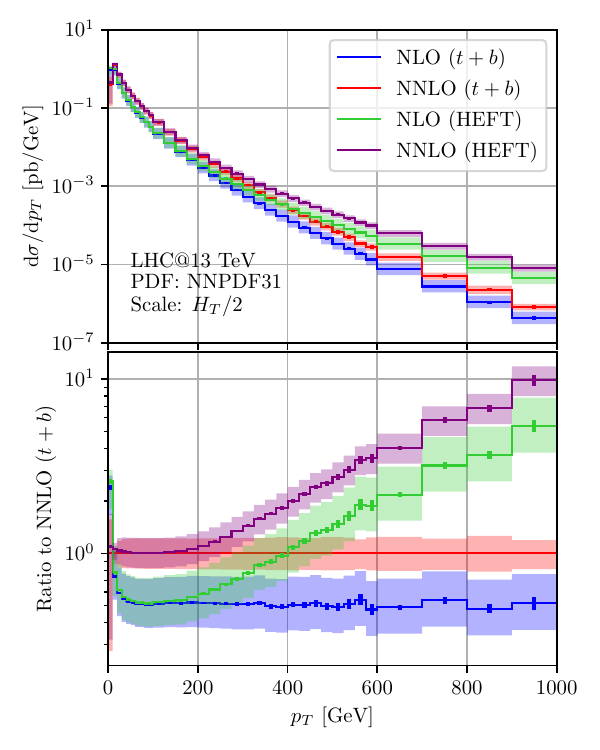}
\end{minipage}
\begin{minipage}[t]{0.49\textwidth}
\includegraphics[width=\textwidth]{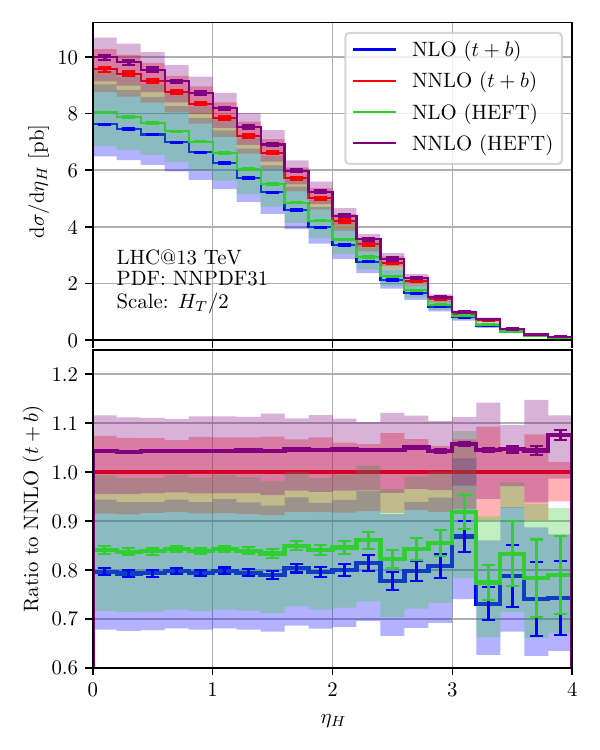}
\end{minipage}
\caption{Higgs-$p_T$ (left) and Higgs-rapidity (right) distribution for the gluon-gluon fusion channel. Compared are 5FS results obtained in HEFT with those including finite-quark-mass effects in the square of the top-quark-loop amplitude and in the interference of the top-quark-loop and bottom-quark-loop amplitudes. The bottom-quark mass is renormalised in the $\overline{\text{MS}}$ scheme, while the top-quark mass is renormalised in the OS scheme.}
\label{fig:pT_yH}
\end{figure}

In figs.~\ref{fig:pT_yH}, we compare distributions in HEFT against the cross sections in full QCD\footnote{Contributions without top quark couplings are excluded, since they are negligible. We also neglect couplings to quarks lighter than the bottom quark.}. As is well-known, the $p_T$ distribution is very sensitive to the inclusion of finite-quark-mass effects, as the large-$p_T$ tail has a different scaling behaviour in HEFT. This can be understood from a simple dimensional analysis. At very high $p_T$, the only relevant mass scale is $p_T$ itself. $\mathrm{d}\sigma/\mathrm{d}p_T^2$ has mass dimension $-4$, ergo in full QCD the tail must scale as 
\begin{equation}
\frac{\mathrm{d}\sigma}{\mathrm{d}p_T^2} \sim  p_T^{-4}.
\end{equation} 
But in HEFT the cross section must always be proportional to $1/v^2$, where $v$ is the vacuum expectation value, because the Higgs-gluon coupling in HEFT always gives a factor of $1/v$. Therefore the large-$p_T$ tail in HEFT scales as 
\begin{equation} 
\frac{\mathrm{d}\sigma}{\mathrm{d}p_T^2} \sim  p_T^{-2} v^{-2}
\end{equation}
and is hence less suppressed. 
Scale-uncertainty bands at large $p_T$ show some slight tension between NLO and NNLO, suggesting that the standard central scale choice \eqref{eq:scale} might not be optimal for this process.

For the Higgs-rapidity distribution in fig.~\ref{fig:pT_yH} we also observe that the scale-uncertainty bands of the different perturbative orders do not overlap. For both full QCD and HEFT, the difference between NLO and NNLO amounts to a rapidity independent upwards shift of about 20\%. Similarly, finite quark masses also only shift the distribution, this time by around $-4\%$, an effect which is directly derived from the interference contribution in fig.~\ref{fig:pT_yH_comparison}.

\section{Conclusions}
\label{sec:sec4}
In this work, we studied the impact of using variable flavour number and different mass-renormalisation schemes on the Higgs production cross section in the gluon-gluon fusion channel. We found that the two different flavour schemes are in excellent agreement for the top-bottom interference contribution, finally justifying our treatment of the bottom-quark mass in the 5FS. The total cross section differs by around 4\% between different FSs and is consistent within scale uncertainties. Generally speaking, it is advisable and more common to use the 5FS for LHC phenomenology because it automatically resums logarithms of the form $\log(m_b^2/s)$. Our computation in two different renormalisation schemes showed that renormalising the bottom-quark mass in the $\overline{\text{MS}}$ scheme results in significantly smaller scale uncertainties and displays a better perturbative convergence. The renormalisation of the top-quark mass on the other hand does not impact the cross section significantly. We therefore recommend using the 5FS, an $\overline{\text{MS}}$-renormalised bottom mass and an OS-renormalised top mass going forward. The cross section for this set-up reads\footnote{The $\mathrm{N}^3\mathrm{LO}$ HEFT cross section was computed using \texttt{SusHi} \cite{Harlander:2012pb, Harlander:2016hcx}.}
\begin{equation}
\boxed{\sigma_{ggH} = 48.81(1)^{+0.65}_{-2.02} (\text{N}^3\text{LO HEFT}) - 0.16^{+0.13}_{-0.03}(\text{NNLO } t) - 1.74(2)^{+0.13}_{-0.03} (\text{NNLO } t\times b )~\text{pb}.}
\end{equation}

We further provided new results for the differential cross section distribution for the Higgs rapidity, which will be useful in upcoming LHC analyses, especially for determining the Yukawa couplings of the lighter quarks.  

\begin{acknowledgments}
We would like to thank Jonas Lindert for providing us with raw data from ref.~\cite{Caola:2018zye}. This work was supported by the Deutsche Forschungsgemeinschaft (DFG) under grant 396021762 -
TRR 257: Particle Physics Phenomenology after the Higgs Discovery, and grant 400140256 - GRK
2497: The Physics of the Heaviest Particles at the LHC. The authors gratefully acknowledge the computing time provided to them at the NHR Center NHR4CES at RWTH Aachen University (project number p0020025). This is funded by the Federal Ministry of Education and Research, and the state governments participating on the basis of the resolutions of the GWK for national high performance computing at universities (\url{www.nhr-verein.de/unsere-partner}).
\end{acknowledgments}

\appendix
\section{Comparison to Higgs+jet analyses}
\label{sec:appendix1}
In this appendix, we compare our results for the Higgs-$p_T$ spectrum with existing analyses of the Higgs+jet process.

\begin{figure}
\centering
\includegraphics[scale=0.7]{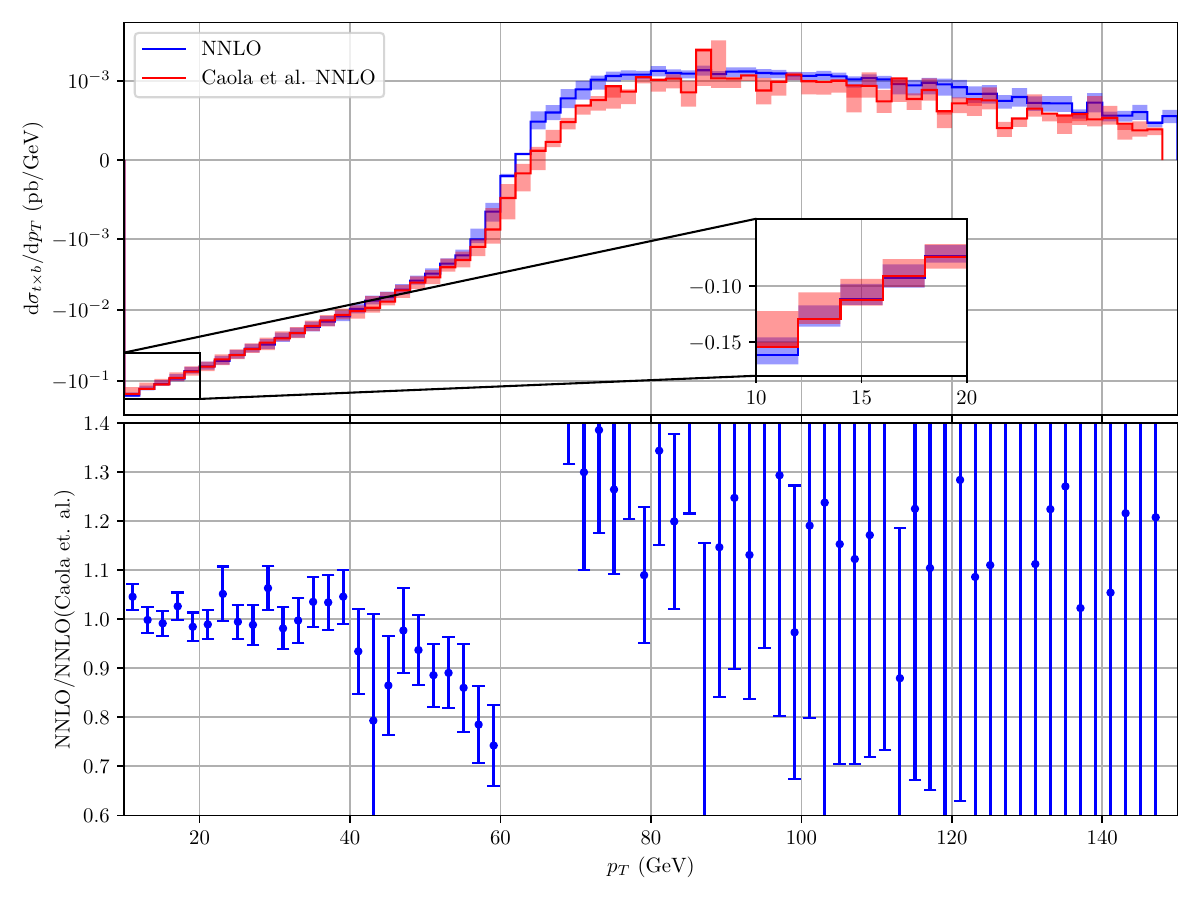}
\caption{Top-bottom interference contribution to the Higgs production cross section. Displayed are our results for the Higgs-$p_T$ spectrum and the results presented in ref.~\cite{Caola:2018zye}, obtained from private communication. Our data was computed with $m_t = 173.05$~GeV, $m_b = 4.78$~GeV and $m_H = 125$~GeV, whereas the authors of ref.~\cite{Caola:2018zye} used $m_t = 173.2$~GeV, $m_b = 4.75$~GeV and the same Higgs mass. For the sake of this comparison we used the same PDF and scale choice as the authors of ref.~\cite{Caola:2018zye}, namely \texttt{PDF4LHC15\_nlo\_30} and $\mu = \frac{1}{2} \sqrt{m_H^2 + p^2_{T}}$. Transparent bands indicate scale uncertainties, whereas the error bars in the lower plot indicate the Monte Carlo uncertainties dominated by the uncertainties of ref.~\cite{Caola:2018zye}.}
\label{fig:Caola_comparison}
\end{figure}

In the OS scheme, we tested our results for the Higgs-$p_T$ spectrum against the fixed-order results presented in ref.~\cite{Caola:2018zye}. In that reference, the authors computed the two-loop real-virtual corrections under the assumption of an infinitely heavy top quark and a nearly massless bottom quark. Fig.~\ref{fig:Caola_comparison} shows that these approximations perform well for $p_T < 40$~GeV, but result in an error of approximately 20\% above this threshold. The nearly massless bottom quark approximation is expected to be valid if $m_b \ll m_H, p_T$, suggesting that most of the error arises from the HTL approximation. Additionally, we observe that the upwards scale variations in ref.~\cite{Caola:2018zye} are significantly larger in the first two bins ($p_T<14$~GeV). After reaching out to the authors, we identified that this discrepancy is due to one specific scale variation ($\mu_R/\mu = 1, \mu_F/\mu = 2$), but we have not found any errors on our end that would account for this difference.

\begin{figure}
\centering
\includegraphics[scale=0.7]{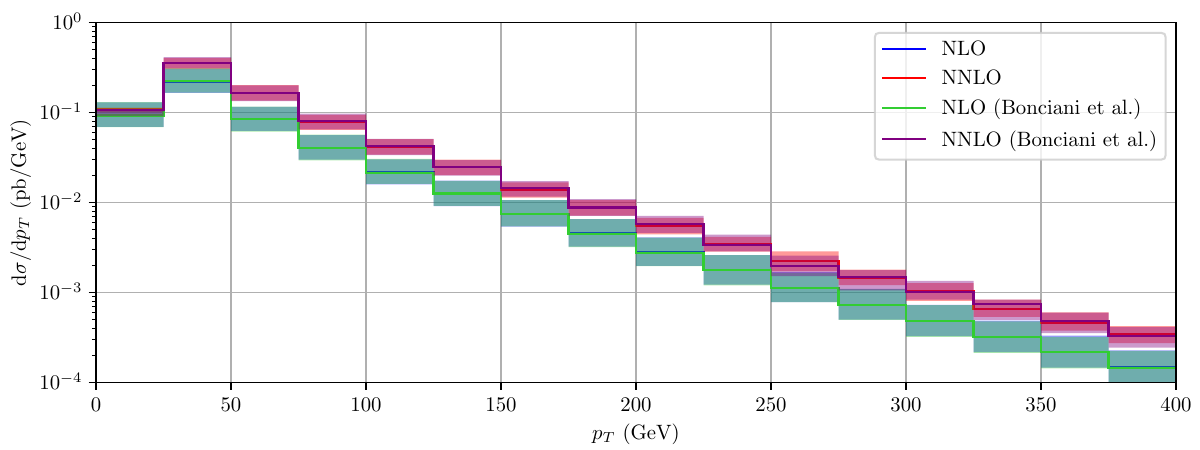}
\caption{Result for the differential cross section of gluon-gluon fusion and comparison to ref.~\cite{Bonciani:2022jmb}. The result is computed with finite top- and bottom-quark mass, where we neglect the cross-section contribution that does not contain any top-Higgs coupling at NNLO. For the comparison we use the same computational set-up as \cite{Bonciani:2022jmb}, i.e.~mass values of $\overline{m}_t = 163.4$~GeV, $\overline{m}_b = 4.18$~GeV, $m_H = 125.25$~GeV, while the running of the mass is computed at two-loop accuracy. We apply the anti-$k_T$ jet algorithm with $R=0.4$ and use the central scale of $\mu = \frac{1}{2} \left( \sqrt{m_H^2 + p_T^2} + \sum_i p_{i,T} \right)$, where the sum runs over all jet momenta. We apply a cut on the transverse momentum of the jets, requiring that at least one of the jet transverse momenta satisfies $p_{i,T}>20$~GeV. We use the \texttt{NNPDF40\_nlo\_as\_01180} PDF set.}
\label{fig:Bonciani_comparison}
\end{figure}

We also compared our $\overline{\mathrm{MS}}$-renormalised results, i.e.\ with both the top- and bottom-quark mass in the $\overline{\mathrm{MS}}$ scheme, for the Higgs-$p_T$ distribution with those in ref.~\cite{Bonciani:2022jmb}. For the real-virtual corrections, we only computed the amplitudes with a closed top-quark loop for scales up to $\mu_R = 125\ \mathrm{GeV}$. Above this scale, we have to extrapolate outside of our numerical grids. As we can see from fig.~\ref{fig:Bonciani_comparison}, we find excellent agreement between the predictions up to transverse momenta of $400\ \mathrm{GeV}$. Above this scale, we find that the predictions diverge, which is caused by the poor precision of the extrapolation.

\section{Alternative HEFT prescription}
Instead of defining the HEFT framework in terms of the LO $K$-factor (see eq.\ref{eq:HEFT_rescaling}), some studies like ref.~\cite{Anastasiou:2016cez} additionally consider an NLO rescaling prescription defined by
\begin{equation}
\sigma^{\text{N}^n\text{LO}}_{\text{HEFT}} = \frac{\sigma^{\text{NLO}}}{\sigma^{\text{NLO}}_{\text{HTL}}} \sigma_{\text{HTL}}^{\text{N}^n\text{LO}} \approx 1.056 \times \sigma^{\text{N}^n\text{LO}}_{\text{HTL}}.
\end{equation}
The rescaling factor $\sigma^{\text{NLO}}/\sigma_{\text{HTL}}^{\text{NLO}}$ is evaluated at the central scale only, since a scale dependent rescaling factor would result in a renormalisation-scheme dependent HEFT cross section. 

The effect of finite top-quark masses with this HEFT prescription is displayed in tab.~\ref{tab:top-HEFT2}. We see that the effect of the finite top-quark-masses is of similar magnitude---though slightly larger---than in our original rescaling method. Hence, both prescriptions provide accurate approximations for finite top-quark mass effects. 

\begin{table*}[t]
\caption{Effect of the finite top-quark mass on the gluon-gluon fusion cross section for two different computational setups. The results are computed for LHC @ 13 TeV using the \texttt{NNPDF31\_nnlo\_as\_0118} PDF set in the 5FS and the \texttt{NNPDF31\_nnlo\_as\_0118\_nf\_4} PDF set in the 4FS. The central scale is chosen at $\mu_R = \mu_F = m_H/2$. The scale uncertainties are determined with seven-point variation.}
\label{tab:top-HEFT2}
\centering
\begin{tabular}{ccc}
\hline
Order & \multicolumn{2}{c}{$(\sigma_{t} - \sigma_\text{HEFT})$ [pb]} \\
\hline
\hline
\multicolumn{3}{c}{$\sqrt{s}=13$~TeV} \\
\hline
& 5FS & 4FS \\
& $m_t = 173.06$ GeV &  $m_t = 173.06$ GeV \\ 
& & $\overline{m}_b(\overline{m}_b)=4.18$ GeV\\
\hline
LO & $0.13^{+0.04}_{-0.03}$ & $0.12^{+0.03}_{-0.02}$ \\
\hline
NLO & $0^{+0.08}_{-0.17}$ & $0^{+0.05}_{-0.13}$ \\
\hline
NNLO & $0.23^{+0.16}_{-0.07}$ & $0.18^{+0.13}_{-0.06}$ \\
\hline
\end{tabular}
\end{table*}

\newpage
\bibliographystyle{JHEP}
\bibliography{main} 

\end{document}